\documentclass[smallextended]{svjour3}
\smartqed
\usepackage{graphicx}
\usepackage{amsmath}
\usepackage{amssymb}
\newtheorem{Theorem}{Theorem}
\newtheorem{Corollary}{Corollary}
\newtheorem{Lemma}{Lemma}
\newtheorem{Conjecture}{Conjecture}
\newtheorem{Definition}{Definition}

\begin{document}
	
	\title{The Independence of Distinguishability and the Dimension of the System}
	\author{Hao Shu}
	\institute{Hao Shu \at
		College of Mathematics, South China University of Technology, Guangzhou, 510641, P. R. China
		\\
		\email{$Hao\_ B\_ Shu@163.com$}
	}	
	\date{}
	\maketitle
		
	\begin{abstract}	
		The are substantial studies on distinguishabilities, especially local distinguishability, of quantum states. It is shown that a necessary condition of a local distinguishable state set is the total Schmidt rank not larger than the system dimension. However, if we view states in a larger system, the restriction will be invalid. Hence, a nature problem is that can indistinguishable states become distinguishable by viewing them in a larger system without employing extra resources. In this paper, we consider this problem for (perfect or unambiguous) LOCC$_{1}$, PPT and SEP distinguishabilities. We demonstrate that if a set of states is indistinguishable in $\otimes _{k=1}^{K} C^{d _{k}}$, then it is indistinguishable even being viewed in $\otimes _{k=1}^{K} C^{d _{k}+h _{k}}$, where $K, d _{k}\geqslant2, h _{k}\geqslant0$ are integers. This shows that such distinguishabilities are properties of states themselves and independent of the dimension of quantum system. Our result gives the maximal numbers of LOCC$_{1}$ distinguishable states and can be employed to construct a LOCC indistinguishable product basis in general systems. Our result is suitable for general states in general systems. For further discussions, we define the local-global indistinguishable property and present a conjecture.
		\\
	\end{abstract}

	\section{Introduction}

	\qquad In quantum information theory, the distinguishability of states is of central importance. If general POVMs are allowed, states can be distinguished if and only if they are orthogonal\cite{1}. However, in realistic tasks, multipartite states are often shared by separated owners, who can not employ general POVMs. Fortunately, technologies of classical communications have been well-developed and can be employed easily. In spirit of this, distinguishing states by local operators and classical communications (LOCC) becomes available and significant. On the other hand, since the distinguishability via LOCC POVMs implies the distinguishability via SEP POVMs, which implies the distinguishability via PPT POVMs, while PPT and SEP POVMs have more simple properties than LOCC ones, they are also considerable.
	
	There are substantial researches on distinguishabilities, especially LOCC distinguishability. It has been shown that two orthogonal pure states are LOCC distinguishable\cite{1}. An innocent intuition might be that the more entanglement a set has, the harder it can be distinguished by LOCC, which, however, is not true in general. Although entanglement indeed gives bounds to the LOCC distinguishability\cite{3,4}, a LOCC indistinguishable set of nine orthogonal product states in $C^3 \otimes C^3$ exists\cite{2}, which of course has no entanglements. 
	
	The local distinguishability of maximally entangled states might be the most-researched one. In $C^3 \otimes C^3$, three orthogonal maximally entangled states are LOCC distinguishable\cite{5} while there exist three LOCC$_{1}$ indistinguishable orthogonal maximally entangled states in $C^d \otimes C^d$, for $d\geqslant4 $ be even or $d=3k+2$\cite{6}. The result was weakened but extended to all $d\geqslant4 $, namely there are 4 LOCC$_{1}$ indistinguishable orthogonal maximally entangled states in $C^d \otimes C^d$ when $d\geqslant4 $\cite{7}. More results were given for Bell states and generalized Bell states. In $C^2 \otimes C^2$, three Bell states are LOCC indistinguishable\cite{8}, while in $C^d \otimes C^d$ with $d\geqslant3$, three generalized Bell states are LOCC distinguishable\cite{9}. A result shows that if $d$ is a prime and $l(l-1)\leqslant 2d$, then $l$ generalized Bell states are LOCC distinguishable\cite{10}. Note that $d+1$ maximally entangled states in $C^d \otimes C^d$ are LOCC indistinguishable\cite{4}. Strangely, a set of generalized Bell states is LOCC distinguishable by two copies\cite{11}. 
	
	On the other hand, LOCC distinguishability of orthogonal product states is also interesting. It has been shown that an unextendible product basis is LOCC indistinguishable\cite{12} while the LOCC distinguishabllity of a completed product basis is equivalent to its LPCC distinguishability\cite{19}. Constructing LOCC indistinguishable product states is also studied\cite{13,14,15,16,17}. We mention that in $C^3 \otimes C^2$, there are four LOCC$_{1}$ indistinguishable orthogonal product states when Alice goes firstly, while in $C^3 \otimes C^3$, there are five LOCC$_{1}$ indistinguishable orthogonal product states no matter who goes firstly\cite{18}. Other methods include \cite{20,21,22,23}.
	
	As for LOCC indistinguishable sets, auxiliary resources might be employed distinguishing\cite{24,25,26,27,28}. Another scheme is distinguishing with an inconclusive result, known as the unambiguous discrimination\cite{23,29,30,31,32,33,34,35}. Finally, there are works for asymptotically LOCC distinguishability\cite{36,37}.
	
	Instead of considering states in a fixed system as most previous researches, we consider the nature problem that whether a set of indistinguishable states can become distinguishable by viewing them in a larger system without employing other resources. It is considerable for at least three reasons. Firstly, the local distinguishability of states is bounded by the dimension of the system. In a bipartite system, a necessary condition of a local distinguishable set is the total Schmidt rank not larger than the system dimension\cite{4}, which however, can be removed by viewing states in a larger system. Hence, whether the states remain indistinguishable is still suspectable. Secondly, by employing extra resources such as entanglement, a local indistinguishable set may become distinguishable\cite{24,25,26,27,28}, while, an universal resource might only exist in a larger system\cite{27}. This gives a feeling that the local indistinguishability of states might depend on the system. Finally, the distinguishability of points in a Hilbert space could be described by distances. For example, setting the distance of two different points be 1, and otherwise be 0. However, the distance of points may depend on the chosen space. For instance, the usual distance of two diagonal points of a unit square is 2 in the 1-dimensional space consisting of the edges of the square, while it is $\sqrt 2$ in a 2-dimensional space consisting of the plane of the square. Hence, it is worth suspecting that the distinguishability of states may depend on the chosen space.
	
	In this paper, we demonstrate that LOCC$_{1}$, unambiguous LOCC, PPT and SEP distinguishabilities are properties of states themselves, namely independent of the system dimension, by proving that an indistinguishable set in $\otimes _{k=1}^{K} C^{d _{k}}$ remains indistinguishable via  POVMs of the same kind even being viewed in $\otimes _{k=1}^{K} C^{d _{k}+h _{k}}$. Our result solves the problem of searching the maximal number of local distinguishable states once for all and provides a LOCC indistinguishable product basis in general systems. Note that the result is suitable for general states in general systems and both perfect and unambiguous discriminations.
	
	\section{Setting}
	
	\qquad In mathematics, a quantum system shared by $K$ owners, namely partite $A^{(s)}, s=1,2,...,K$, can be described as a Hilbert space $\otimes _{k=1}^{K} C^{d _{k}}$, while a general state can be described as a density operator (positive semi-definited with unit trace). A state set $S$ is LOCC(LOCC$_{1}$, PPT, SEP) distinguishable means that there is a LOCC(LOCC$_{1}$, PPT, SEP) POVM $\left\{\ M _{j}\right \}_{j=1,2,...,J}$ such that for any $j$, $Tr(M_{j}\rho)\neq 0$ for at most one state $\rho$ in $S$.
	
    A LOCC$_{r}$ POVM is described as follow. Partite $A^{(s)}$, $s=1,2,...,K$, provide local measurements, depending on the previously published results, on their partita and publish the results, in order. A round is that every member measures and publishes the result once. A POVM $\left\{\ M _{j}\right \}_{j=1,2,...,J}$ generated by such a procedure after $r$ rounds is defined to be a LOCC$_{r}$ POVM. For example, a $LOCC _{1}$ POVM is given as follow. $A^{(1)}$ provides a POVM $\left\{\ A^{(1)} _{j} \right \}_{j=1,2,...,J}$ on his partita and gets an outcome $j _{1}$. For $s=2,3,...,K$, each $A^{(s)}$ provides a POVM $\left\{\ A^{(s)} _{j _{1},j _{2},...,j _{(s-1)},j} \right \}_{j=1,2,...,J _{j _{1},j _{2},...,j _{(s-1)}}}$ on his partita depending on the classical communications of others measured before him, and gets an outcome $j _{s}$. Hence, the LOCC$_{1}$ POVM is $\left\{\ M _{j _{1},j _{2},...,j _{(K-1)},j _{K}} \right \}_{1\leqslant j _{s}\leqslant J _{j _{1},j _{2},...,j_ {(s-1)}}}$, where $M _{j _{1},j _{2},...,j _{(K-1)},j _{K}}=\otimes _{s=1}^{K} A^{(s)} _{j _{1},j _{2},...,j _{(s-1)},j _{s}}$ and $J _{j _{1},j _{2},...,j_ {(s-1)}}=J$ for $s=1$. On the other hand, $\left\{\ M _{j}\right \}_{j=1,2,...,J}$ is said to be a PPT POVM if every $M _{j}$ is positive semi-definited after a partial transposition, while it is said to be a SEP POVM if every $M _{j}$ can be written as a tensor product of local operators.

    There is a nature embedding from $C^{d}$ to $C^{d+h}$, viewing a state in $\otimes _{k=1}^{K} C^{d _{k}}$ as a state in $\otimes _{k=1}^{K} C^{d _{k}+h _{k}}$. Precisely, a computational basis in $\otimes _{k=1}^{K} C^{d _{k}}$ can be extended to a computational basis in $\otimes _{k=1}^{K} C^{d _{k}+h _{k}}$, by which all operators are written in matrix form. For a density matrix(state) $\rho$ in $\otimes _{k=1}^{K} C^{d _{k}}$, it can be viewed as the density matrix(state)
    $\widetilde{\rho}=
    \begin{pmatrix}
    	\rho & 0\\
    	 0   & 0
    \end{pmatrix}$ in $\otimes _{k=1}^{K} C^{d _{k}+h _{k}}$.

    These views will be employed in the rest of the paper.

	\section{Result}

	\begin{Theorem} 
		Let $\left\{\ \rho _{i} |i=1,2,...,N \right \}$ be a set of states (pure or mixed), written in density matrix form, in $\otimes _{k=1}^{K} C^{d _{k}}$, where N is a finite positive integer. If they are indistinguishable via LOCC$_{1}$(PPT, SEP, global) or unambiguous LOCC (PPT, SEP, global) POVMs, in $\otimes _{k=1}^{K} C^{d _{k}}$, then they are indistinguishable via LOCC$_{1}$(PPT, SEP, global) or unambiguous LOCC (PPT, SEP, global) POVMs, in $\otimes _{k=1}^{K} C^{d _{k}+h _{k}}$ (viewed as $\widetilde{\rho_{i}}=
		\begin{pmatrix}
			\rho _{i} & 0\\
			0        & 0
		\end{pmatrix}$), respectively, where $h_{k}$ are non-negative integers.
	\end{Theorem}
	
	Note that unambiguous LOCC distinguishability is equivalent to unambiguous SEP distinguishability\cite{23}.
	
	The following corollaries, providing maximal number of distinguishable states, generalize results in special systems to general systems and thus somehow show the abilities of the theorem.
	
	For general pure states, since there exist three LOCC$_{1}$ indistinguishable orthogonal states in $C^2 \otimes C^2$, for example, three orthogonal Bell states, Theorem 1 together with the result in \cite{1} imply that:
	
	\begin{Corollary}
		In any non-trivial system, the maximal number T such that any T orthogonal pure states are LOCC$_{1}$ distinguishable is 2.
	\end{Corollary}
	
	For orthogonal product states, four LOCC$_{1}$ indistinguishable states were constructed in $C^3 \otimes C^2$, providing fixed measurement order, while five LOCC$_{1}$ indistinguishable states were constructed in $C^3 \otimes C^3$, providing choice measurement order. Results in bipartite systems also show that three orthogonal product states are LOCC$_{1}$ distinguishable in any order while four orthogonal product states are LOCC$_{1}$ distinguishable in suitable order\cite{18}.
	
	Therefore, as a consequence of Theorem 1, we have:
	
	\begin{Corollary}
		In $C^m \otimes C^n$, the maximal number P such that any P orthogonal product states are LOCC$_{1}$ distinguishable in fixed measurement order is 3, where $m\geqslant 3, n\geqslant 2$, while, the maximal number Q such that any Q orthogonal product states are LOCC$_{1}$ distinguishable in suitable order is 4, where $m, n\geqslant 3$.
	\end{Corollary}
	
	For orthogonal product basis, a LOCC indistinguishable basis in $C^3 \otimes C^3$ was constructed\cite{2}. Assisted with the result in \cite{19}, by using Theorem 1, a LOCC indistinguishable orthogonal product basis in $C^m \otimes C^n$ can be constructed, where $m, n\geqslant 3$.
	
	\begin{Corollary}
		The nine Domino states in \cite{2} together with $|i\rangle|j\rangle$, i=3,4,...,$m-1$, j=3,4,...,$n-1$, form a LOCC indistinguishable completed orthogonal product basis in $C^m \otimes C^n$, where $m, n\geqslant 3$.
	\end{Corollary}
	
	The above basis is LOCC indistinguishable, not only LOCC$_{1}$ indistinguishable. Details are provided as follow.
	
	The nine Domino states (unnormalized) in \cite{2} form an orthogonal product basis in $C^3 \otimes C^3$. They are $|0\rangle |0\pm 1\rangle, |0\pm 1\rangle|2\rangle, |2\rangle |1\pm 2\rangle, |1\pm 2\rangle|0\rangle, |1\rangle|1\rangle$. It is easy to see that in $C^m \otimes C^n$ with $m, n\geqslant 3$, the states together with $|i\rangle|j\rangle, i=3,4,...,m-1, j=3,4,...,n-1$, form a completed orthogonal product basis. We will demonstrate that it is LOCC indistinguishable.
	
    We need a lemma which is an easy corollary of the result in \cite{19}.
	
	\begin{Lemma} An orthogonal product basis in a multipartite system is LOCC distinguishable if and only if it is LOCC$_{1}$ distinguishable.
	\end{Lemma}
	
	By Theorem 1, the above orthogonal basis is LOCC$_{1}$ indistinguishable, since the Domino states are LOCC (and thus LOCC$_{1}$) indistinguishable in $C^3 \otimes C^3$. As we are considering an orthogonal product basis, the above lemma shows that it is LOCC indistinguishable.
	
	The construction can be generalized to multipartite systems, by the tensor product of above basis after normalizing and a normalized orthogonal basis of other partite.

	\section{Proof of the result}

	We only prove the theorem for perfect discriminations. The proofs for unambiguous discriminations are similar.
	
    We will show that the up-left block of a POVM in $\otimes _{k=1}^{K} C^{d _{k}+h _{k}}$ is a POVM in $\otimes _{k=1}^{K} C^{d _{k}}$ of the same kind for LOCC$_{1}$, PPT, SEP or global POVM, while, the condition of distinguishability in a lower dimensional system is the same as in the larger dimensional system, since the low-right block has trace 0.
    	
	\begin{Lemma}
		Let $\left\{\ M_{j} \right \}_{j=1,2,...,J}$ be a (LOCC$_{1}$, PPT, SEP, global) POVM of $\otimes _{k=1}^{K} C^{d _{k}+h _{k}}$ such that
		$M _{j}=\begin{pmatrix}
			M_{j1} & M_{j2}\\
			M_{j3} & M_{j4}
		\end{pmatrix}$
		written in block form, where $M _{j1}$ is a $(\prod _{k=1}^{K}d _{k})\times(\prod _{k=1}^{K}d _{k})$ matrix. Then $\left\{\ M_{j1} \right \}_{j=1,2,...,J}$ is a (LOCC$_{1}$, PPT, SEP, global) POVM of $\otimes _{k=1}^{K} C^{d _{k}}$.
	\end{Lemma}
	
	\noindent \textbf{Proof:}
	
	For a POVM $\left\{\ M_{j} \right \}_{j=1,2,...,J}$ in $\otimes _{k=1}^{K} C^{d _{k}+h _{k}}$, $M_{j}$ can be written as 
	
	$M _{j}=\sum _{i} (\otimes _{s=1}^{K}A^{(s)}_{ji})$, where the sum is finite. Write
	$A^{(s)} _{ji}=\begin{pmatrix}
		A^{(s)}_{ji1} & A^{(s)}_{ji2}\\
		A^{(s)}_{ji3} & A^{(s)}_{ji4}
	\end{pmatrix}$,
    $M _{j}=\begin{pmatrix}
	M_{j1} & M_{j2}\\
	M_{j3} & M_{j4}
\end{pmatrix}$
	in block form, where $A^{(s)}_{ji1}$ is a $d _{s}\times d _{s}$ matrix, $M_{j1}$ is a $\prod _{s=1}^{K}d _{s}\times \prod _{s=1}^{K}d _{s}$ matrix. Then $M _{j1}=\sum _{i} \otimes _{s=1}^{K}(A^{(s)} _{ji1})$. Since $\left\{\ M _{j} \right \}_{j=1,2,...,J}$ is a POVM, $\sum _{j}M _{j}=I_{\prod _{s=1}^{K}(d _{s}+h _{s})}$ while $M _{j}$ is positive semi-definited. Therefore, $\sum _{j}M_{j1}=I_{\prod _{s=1}^{K} d _{s}}$ while $M _{j1}$ is positive semi-definited. Hence, $\left\{\ M _{j1} \right \}_{j=1,2,...,J}$ is a POVM.	
	
	$\left\{\ M _{j} \right \}_{j=1,2,...,J}$ is a $PPT$ POVM means that for every j, $M _{j}$ is positive semi-definited after a partial transposition. Without loss generality, for j, assume that the partial transposition is on $A^{(1)}$. Hence, $\sum _{i} [(A^{(1)} _{ji})^{T}\otimes \otimes _{s=2}^{K}(A^{(s)} _{ji})]$ is positive semi-definited and thus, $\sum _{i} [(A^{(1)} _{ji1})^{T}\otimes \otimes _{s=2}^{K}(A^{(s)} _{ji1})]$ is positive semi-definited, which implies that $M _{j1}$ is a $PPT$ operator.
	
    $\left\{\ M_{j} \right \}_{j=1,2,...,J}$ is a $SEP$ POVM means that $M_{j}$ can be written as 
    
    \noindent $M _{j}=\otimes _{s=1}^{K}A^{(s)} _{j}$, for every $j$. Hence, $M _{j1}=\otimes _{s=1}^{K}A^{(s)} _{j1}$ is separable, which implies that $\left\{\ M _{j1} \right \}_{j=1,2,...,J}$ is a $SEP$ POVM of $\otimes _{s=1}^{K} C^{d _{k}}$, where similar to above,
	$A^{(s)} _{j}=\begin{pmatrix}
		A^{(s)}_{j1} & A^{(s)}_{j2}\\
		A^{(s)}_{j3} & A^{(s)}_{j4}
	\end{pmatrix}$.

    On the other hand, as in the setting section, let $\left\{\ M _{j _{1},j _{2},...,j _{(K-1)},j _{K}} \right \}_{1\leqslant j _{s}\leqslant J _{j _{1},j _{2},...,j_ {(s-1)}}}$ be a $LOCC_{1}$ POVM, where $M _{j _{1},j _{2},...,j _{(K-1)},j _{K}}=\otimes _{s=1}^{K} A^{(s)} _{j _{1},j _{2},...,j _{(s-1)},j _{s}}$. Write

  \noindent $M _{j _{1},j _{2},...,j _{(K-1)},j _{K}}=
	\begin{pmatrix}
		M _{j _{1},j _{2},...,j _{(K-1)},j _{K}1} &
		M _{j _{1},j _{2},...,j _{(K-1)},j _{K}2}\\
		M _{j _{1},j _{2},...,j _{(K-1)},j _{K}3} &
		M _{j _{1},j _{2},...,j _{(K-1)},j _{K}4}
	\end{pmatrix}$,

  \noindent $A^{(s)} _{j _{1},j _{2},...,j _{(s-1)},j _{s}}=\begin{pmatrix}
	A^{(s)}_{j _{1},j _{2},...,j _{(s-1)},j _{s}1} & A^{(s)}_{j _{1},j _{2},...,j _{(s-1)},j _{s}2}\\
	A^{(s)}_{j _{1},j _{2},...,j _{(s-1)},j _{s}3} & A^{(s)}_{j _{1},j _{2},...,j _{(s-1)},j _{s}4}
\end{pmatrix}$, 

  \noindent where $M_{j _{1},j _{2},...,j _{(K-1)},j _{K}1}$ is a $\prod _{s=1}^{K}d _{s}\times \prod _{s=1}^{K}d _{s}$ matrix, $A^{(s)}_{j _{1},j _{2},...,j _{(s-1)},j _{s}1}$ is a $d _{s}\times d _{s}$ matrix. Hence, $M _{j _{1},j _{2},...,j _{(K-1)},j _{K}1}=\otimes _{s=1}^{K} A^{(s)} _{j _{1},j _{2},...,j _{(s-1)},j _{s}1}$. Since $\left\{\ A^{(s)} _{j _{1},j _{2},...,j _{(s-1)},j _{s}} \right \}_{j _{s}=1,2,...,J _{j _{1},j _{2},...,j_ {(s-1)}}}$ is a local POVM of partita $A^{(s)}$, 
  
  \noindent $\left\{\ A^{(s)} _{j _{1},j _{2},...,j _{(s-1)},j _{s}1} \right \}_{j _{s}=1,2,...,J _{j_{1},j_{2},...,j_{(s-1)}}}$ is a local POVM of $A^{(s)}$. Therefore, $\left\{\ M _{j _{1},j _{2},...,j _{(K-1)},j _{K}1} \right \}_{1\leqslant j _{s}\leqslant J _{j _{1},j _{2},...,j_ {(s-1)}}}$ is a LOCC$_{1}$ POVM.
    \\

	\noindent \textbf{Proof of Theorem 1}:
	
	Let $\left\{\ \rho _{i} |i=1,2,...,N \right \}$ be a set of orthogonal states in $\otimes _{s=1}^{K} C^{d _{s}}$. Assume that it is distinguishable via $LOCC _{1}$(PPT, SEP, global) POVM $\left\{\ M _{j} \right \}_{j=1,2,...,J}$ in $\otimes _{s=1}^{K} C^{d _{s}+h _{s}}$, namely, for every $j$, $Tr(M _{j}\widetilde{\rho _{i}})\neq 0$ for at most one i. Let us prove that it is distinguishable via $LOCC _{1}$(PPT, SEP, global) POVM in $\otimes _{s=1}^{K} C^{d _{s}}$.
	
	Write
	$M _{j}=\begin{pmatrix}
		M_{j1} & M_{j2}\\
		M_{j3} & M_{j4}
	\end{pmatrix}$
	in block form, where $M_ {j1}$ is a $\prod _{s=1}^{K}d _{s}\times \prod _{s=1}^{K}d _{s}$ matrix. Since $\left\{\ M _{j} \right \}_{j=1,2,...,J}$ is a $LOCC _{1}$(PPT, SEP, global) POVM, $\left\{\ M _{j1} \right \}_{j=1,2,...,J}$ is a $LOCC _{1}$(PPT, SEP, global) POVM of $\otimes _{s=1}^{K} C^{d _{s}}$, by the above Lemma.
	
	Note that $Tr(M _{j1}{\rho _{i}})\neq 0$ implies that $Tr(M _{j}\widetilde{\rho _{i}})\neq 0$, since that they are equal. Hence, at most one $\rho _{i}$ satisfies $Tr(M _{j1}{\rho _{i}})\neq 0$, which means that the states are $LOCC _{1}$(PPT, SEP, global) distinguishable via POVM $\left\{\ M _{j1} \right \}_{j=1,2,...,J}$ in $\otimes _{s=1}^{K} C^{d _{s}}$.
	$\hfill\blacksquare$
	
	\section{Discussion}
		
	\qquad The result of other indistinguishabilities may also hold. However, the method in this paper may not work. For example, the up-left block of a LOCC(LPCC) POVM in $\otimes _{k=1}^{K} C^{d _{k}+h _{k}}$ may not be a LOCC(LPCC) POVM in $\otimes _{k=1}^{K} C^{d _{k}}$. The reason may relate to that measuring a state in a larger dimensional system, the collapsing state may not in the ordinary lower dimensional system. We construct a projective POVM in $C^4 \otimes C^4$, which is not projective when only looking at the up-left $3 \times 3$ block as follow.
	
		\begin{equation*}
			P_{1}=
			\begin{pmatrix}
				\quad\frac{1}{4} & \quad\frac{1}{4} & \quad\frac{1}{4} & \quad\frac{1}{4}\\
				\quad\frac{1}{4} & \quad\frac{1}{4} & \quad\frac{1}{4} & \quad\frac{1}{4}\\
				\quad\frac{1}{4} & \quad\frac{1}{4} & \quad\frac{1}{4} & \quad\frac{1}{4}\\
				\quad\frac{1}{4} & \quad\frac{1}{4} & \quad\frac{1}{4} & \quad\frac{1}{4}
			\end{pmatrix}
			P_{2}=
			\begin{pmatrix}
				\quad\frac{1}{4} & -\frac{1}{4} & \quad\frac{1}{4} & -\frac{1}{4}\\
				-\frac{1}{4} & \quad\frac{1}{4} & -\frac{1}{4} & \quad\frac{1}{4}\\
				\quad\frac{1}{4} & -\frac{1}{4} & \quad\frac{1}{4} & -\frac{1}{4}\\
				-\frac{1}{4} & \quad\frac{1}{4} & -\frac{1}{4} & \quad\frac{1}{4}
			\end{pmatrix}
		\end{equation*}
			
		\begin{equation*}
			P_{3}=
			\begin{pmatrix}
				\quad\frac{1}{4} & \quad\frac{1}{4} & -\frac{1}{4} & -\frac{1}{4}\\
				\quad\frac{1}{4} & \quad\frac{1}{4} & -\frac{1}{4} & -\frac{1}{4}\\
				-\frac{1}{4} & -\frac{1}{4} & \quad\frac{1}{4} & \quad\frac{1}{4}\\
				-\frac{1}{4} & -\frac{1}{4} & \quad\frac{1}{4} & \quad\frac{1}{4}
			\end{pmatrix}	
			P_{4}=
			\begin{pmatrix}
				\quad\frac{1}{4} & -\frac{1}{4} & -\frac{1}{4} & \quad\frac{1}{4}\\
				-\frac{1}{4} & \quad\frac{1}{4} & \quad\frac{1}{4} & -\frac{1}{4}\\
				-\frac{1}{4} & \quad\frac{1}{4} & \quad\frac{1}{4} & -\frac{1}{4}\\
				\quad\frac{1}{4} & -\frac{1}{4} & -\frac{1}{4} & \quad\frac{1}{4}
			\end{pmatrix}
		\end{equation*}
	
	However, it will not be surprising if the result can be extended to other cases. Note that for a completed or unextendible product basis, the result holds for LOCC distinguishability\cite{19,36}. Hence, we have the following definitions and conjecture:
	
	\begin{Definition}\textbf{(Local-global indistinguishable property)}
		
		Let $\left\{\ d_{k} \right \}_{k=1,2,...,N}$ be integers with $d_{k}\geq 2$. For a state set S in $\otimes _{k=1}^{K} C^{d _{k}}$ and kinds of indistinguishabilities M, $M'$,
		
		(1) Define that S satisfies the $M\rightarrow M'$ local-global indistinguishable property, if S is indistinguishable via M in $\otimes _{k=1}^{K} C^{d _{k}}$ implies that S is indistinguishable via $M'$ in $\otimes _{k=1}^{K} C^{d _{k}+ h_{k}}$, where $h _{k}$ are any non-negative integers.	
		
		(2) If for any state set S in $\otimes _{k=1}^{K} C^{d _{k}}$, S satisfies the $M\rightarrow M'$ local-global indistinguishable property, then $\otimes _{k=1}^{K} C^{d _{k}}$ is said to be satisfying the $M\rightarrow M'$ local-global indistinguishable property.	
		
		(3) If for any state set S in any system, it satisfies the $M\rightarrow M'$ local-global indistinguishable property, then $M\rightarrow M'$ is said to be satisfying the local-global indistinguishable property.
	\end{Definition}
	
	\begin{Conjecture}
		If M=(perfect or unambiguous) LOCC(LPCC, LOCC$_{r}$, LPCC$_{r}$, PPT, SEP, global, projective) distinguishability, then $M\rightarrow M$ satisfies the local-global indistinguishable property.
	\end{Conjecture}
		
	This gives a framework of local-global indistinguishability, which states the independence of state distinguishability and the dimension of the system. Now, Theorem 1 can be restated as: For M=$LOCC _{1}$ (unambiguous LOCC, (perfect or unambiguous) PPT, SEP, global) distinguishability, the conjecture is true.
	
	\section{Conclusion}

	\qquad In this paper, we consider the nature problem that whether the indistinguishability of states depends on the dimension of the system. We demonstrate that LOCC$_{1}$, PPT and SEP indistinguishabilities, both perfect and unambiguous, are properties of states themselves and independent of the dimensional choice. More exactly, we show that if states are LOCC$_{1}$ (or unambiguous LOCC, (perfect or unambiguous) PPT, SEP) indistinguishable in a lower dimensional system, then they are LOCC$_{1}$ (or unambiguous LOCC, (perfect or unambiguous) PPT, SEP) indistinguishable in a dimensional extended system. The result is true for both bipartite and multipartite systems and for both pure and mixed states.
	
	Assisted with previous results, Theorem 1 gives the maximal numbers of local distinguishable states and can be employed to construct a LOCC indistinguishable orthogonal product basis in general systems, except for one or two small dimensional ones. Note that the corollaries are even suitable for multipartite systems.
	
	For further discussions, we define the local-global indistinguishable property and present a conjecture. Both proving the validity and searching counter-examples for it could be interesting.
	
	\section*{Acknowledgments}

	We wish to acknowledge professor Zhu-Jun Zheng who gave advices and helped to check the paper.
	
	\section*{Declarations}
	
	\subsection*{\textbf{Funding}} No funding.
	
	\subsection*{\textbf{Conflicts of interest/Competing interests}} The author declares there are no conflicts of interest.
	
	\subsection*{\textbf{Availability of data and material}} All data supporting the finding can be found in the article.
	
	\subsection*{\textbf{Code availability}} Not applicable.
	
	\subsection*{\textbf{Authors' contributions}} The article has a unique author.

		\end{document}